\documentclass{article}

\usepackage{graphicx}  
\usepackage{amsmath}   
\usepackage[compress]{cite}
\usepackage{amssymb}   
\usepackage{bm} 

\usepackage{mathrsfs}

\def\eq#1{{Eq.~(\ref{#1})}}

\def\gl#1#2{{g_{#1#2}}}

\def\half{{\frac{1}{2}}}

\title{What drives the time evolution of the spacetime geometry?}

\author{T. Padmanabhan\\
IUCAA, Post Bag 4, Ganeshkhind,
 Pune - 411 007, India.\\
email: paddy@iucaa.ernet.in}

\date{ }

\begin{document}

\maketitle

\begin{abstract}

   I  show that in a general, evolving spacetime,  the rate of change of gravitational momentum is related to the difference between the number of   degrees of freedom in the bulk  and the boundary of a region. This expresses the  gravitational field equation in the thermodynamic language (in contrast with the  conventional geometric language) which is the natural description, if gravity is an emergent phenomenon.
In all static spacetimes, the  number of  degrees of freedom in the boundary is equal to the number of  degrees of freedom in the bulk; i.e., these spacetimes maintain holographic equipartition. \textit{It is the departure from  holographic equipartition that drives the time evolution of the spacetime.} This result, which is equivalent to Einstein's equations, provides an elegant, holographic, description of spacetime dynamics.
\end{abstract}

 The  \textit{mathematical} answer to the question in the title is given by  Einstein's equation $G^a_b=8\pi T^a_b$, which  determines the  metric in terms of the matter source.
 But what does Einstein's equation mean \textit{physically}? I will show that one can provide an elegant, holographic answer in terms of an alternative equation:
  \begin{equation}
 \int_\mathcal{V}\frac{d^3x}{8\pi}h_{ab}\pounds_\xi p^{ab}  = \frac{1}{2} k_B T_{\rm avg} ( N_{\rm bulk} - N_{\rm sur})
\label{start}
\end{equation} 
Here, $h_{ab}$ is the induced metric on the $t=$ constant surface, $p^{ab}$ is its conjugate momentum and $\xi^a=Nu^a$ is the proper-time evolution vector corresponding to observers moving with four-velocity $u_a  = - N \nabla_a t$ which is the normal to the $t=$ constant surface. The $N_{\rm sur}$ and $N_{\rm bulk}$ are the degrees of freedom in the surface $\partial\mathcal{V}$ and bulk $\mathcal{V}$ of a 3-dimensional region  and $T_{\rm avg}$ is the average Davies-Unruh temperature \cite{du}  of the boundary. The left hand side is the time rate of change of gravitational momentum which is driven by the departure from holographic equipartition, indicated by a non-zero value for $(N_{\rm bulk} - N_{\rm sur})$. The time evolution will disappear when $N_{\rm sur} = N_{\rm bulk}$ and, in fact, all static geometries obey this condition of holographic equipartition.  The validity of \eq{start} for all observers (i.e., foliations) ensures the validity of Einstein's equations; thus, \eq{start} carries  the same physical content as the gravitational field equations. In short, \textit{holographic equipartition dictates the evolution of spacetime geometry}.

I will now describe how this result arises \cite{grtd}.  
Several recent investigations  suggest that the  gravitational field equations should have the same status as the equations of elasticity or fluid mechanics (for reviews, see e.g., \cite{a8}). 
 This connection becomes most apparent when we use  $f^{ab}\equiv \sqrt{-g}g^{ab}$ as the dynamical variable 
(instead of the usual $\gl ab$)
and the corresponding canonical momenta  $N^c_{ab}$ defined by:
\begin{equation} 
N^{a}_{bc} \equiv -\Gamma^{a}_{bc}+\frac{1}{2}(\Gamma^d_{bd}\delta^{a}_{c}+\Gamma^d_{cd}\delta^{a}_{b}) 
\label{NGamma}                       
\end{equation}  
The variations of these \textit{dynamical} variables ($f^{ab} \delta N^c_{ab}$, $ N^c_{ab} \delta f^{ab}$) and the variations of the \textit{thermodynamic} variables ($S\delta T, T\delta S$) exhibit an one-to-one correspondence \cite{KBP} when evaluated on the null surfaces.\footnote{I use the (-- + + +) signature and units with $k_B=1,c=1,\hbar=1,16\pi G=1$, so that Einstein's equations reduce to $2G_{ab}=T_{ab}$. The Latin letters run through 0-3 while the Greek letters take values 1-3.}
  
  It also turns out that a  similar combination $f^{ab} \pounds_v N^i_{ab}$ occurs  in the expression for the conserved current associated with a vector field $v^a$.
If we separate the derivative $\nabla_k v_j$ of  any vector field $v^j$   into the symmetric and anti-symmetric parts by 
$\nabla^{(j} v^{k)} \equiv S^{jk}$ and $\nabla^{[j} v^{k]} \equiv J^{jk}$, then  the anti-symmetric part $J^{lm}$  immediately gives us a conserved current $J^i\equiv \nabla_k J^{ik}$; in other words, from every vector field $v^k$ in the spacetime, we can obtain a conserved current, rather trivially.   A more useful form for this current can be found as follows:
From the Lie derivative of the connection
$
\pounds_v\Gamma^a_{bc}=\nabla_b \nabla_c v^a+R^a_{\phantom{a}cmb}v^m
$,
we  obtain, on using \eq{NGamma}, the relation: 
$
g^{bc}\pounds_vN^a_{bc} =\nabla_bJ^{ab}-2R^a_bv^b.
$
This gives  the explicit form of the conserved current
\begin{equation}
 J^a[v] = \nabla_b J^{ab} [v] = 2 R^a_b v^b + g^{ij} \pounds_v N^a_{ij}
\label{noe}
\end{equation}
In fact, this \textit{is} indeed the standard Noether current associated with  $v^a$ --- which we have now derived \textit{without mentioning the action principle for gravity or its diffeomorphism invariance!}.

While \eq{noe} associates a conserved current (and charge) with \textit{any} vector field, the one related to the vector describing the time evolution  is of special interest. The vector $\xi^a= Nu^a$ measures the  proper-time lapse corresponding to the normal $u_a  = - N \nabla_a t$ to the  $t=$ constant surfaces in any spacetime. (In static spacetimes,  we can choose $\xi^a$ to be the timelike Killing vector.)   An elementary calculation shows \cite{grtd} that the Noether charge associated with $\xi^a$ has a simple form which --- as we shall see --- admits a direct thermodynamic meaning.
We find that:
\begin{equation}
u_aJ^a(\xi)=2D_\alpha (Na^\alpha)
\label{fin1}
\end{equation} 
where $a^i\equiv u^j\nabla_ju^i$ is the acceleration and   $D_\alpha$ is the covariant derivative on the $t=$ constant surface. 
The acceleration $a_i$ has the explicit form
$
 Na_i=Nu^l\nabla_lu_i=h^j_i\nabla_jN.
$
Integrating \eq{fin1}  over $\sqrt{h}d^3x$ we obtain the total Noether charge contained inside a volume, which is just the flux of the acceleration! Adding  the correct proportionality constant (taking $\hbar=1,c=1,k_B=1,G=L_P^2$), we get:
 \begin{equation}
\int_\mathcal{V}\sqrt{h}\,d^3x\ u_aJ^a[\xi]=\int_\mathcal{V}d\Sigma_aJ^a[\xi]=
\int_{\partial\mathcal{V}}\frac{\sqrt{\sigma}\, d^2x}{8\pi L_P^2} (Nr_\alpha a^\alpha)
\label{flux2}
\end{equation} 
This result is valid 
for any 3-dimensional region $\mathcal{V}$ in any spacetime.

We now choose the boundary to be a $N(t,\mathbf{x})=$  constant surface within the $t=$ constant surface. The $r_\alpha$ is then the  normal to  the $N(t,\mathbf{x})=$ constant surface within the $t=$ constant surface. So one can write $r_\alpha\propto D_\alpha N$ or 
$r_i\propto h^j_i\nabla_jN$ where $h^i_j=\delta^i_j+u^iu_j$ is the projection tensor to the  $t=$ constant surface.  Because $Na_i=h^j_i\nabla_jN$, it follows that $r_i$ and $a_i$ are in the same direction even in the most general (non-static, $N_\alpha\neq0$) spacetime geometry.  This gives $Nr_\alpha a^\alpha=Na=(h^{ij}\nabla_iN\nabla_jN)^{1/2}$.
 Therefore, when the boundary is a surface with $N=$ constant (which generalizes of the notion of an equipotential surface), we can interpret $T_{\rm loc}=Nr_\alpha a^\alpha/2\pi=Na/2\pi$ as the (Tolman redshifted) local Davies-Unruh temperature \cite{du} perceived by the observers with four-velocity $u_a=-N\delta_a^0$. These observers are moving normal to the $t=$constant hypersurfaces with the acceleration $a$ with respect to the local freely-falling observers. The  vacuum state in the freely-falling frame will appear as a thermal state with temperature $T_{\rm loc}=Na/2\pi$ to these observers. Hence we can write:
\begin{equation}
2\int_\mathcal{V}\sqrt{h}\, d^3x\ u_aJ^a[\xi]=
\int_{\partial\mathcal{V}}\frac{\sqrt{\sigma}\, d^2x}{L_P^2} \left(\frac{1}{2}T_{\rm loc}\right)
\label{ib1}
\end{equation}
This equation relates (twice) the Noether charge contained in a $N=$ constant surface  to the  equipartition energy of the surface, when we attribute one  degree of freedom to each cell of Planck area $L_P^2$.
An equivalent interpretation arises, if we think of
$s=\sqrt{\sigma}/4L_P^2$ as the analogue of the entropy density. We then get, directly from \eq{flux2}, the relation: 
\begin{equation}
\int_\mathcal{V}\sqrt{h}\ d^3x\ u_aJ^a[\xi]
=\int_{\partial\mathcal{V}}d^2x\ Ts
\label{flux3}
\end{equation}
where the right-hand-side is the integral of the heat (enthalpy)  density ($TS/A$) of the boundary surface. 
Thus, the Noether charge (for the time-development vector) contained in a region of space bounded by an $N(t,\mathbf{x})=$ constant surface, is equal to the surface heat content. This delightfully simple interpretation is valid in \textit{the most general} context without any assumptions like static nature, existence of Killing vectors, asymptotic behaviour, etc.

Incidentally, the
 factor 2 on the left hand side of \eq{ib1}, also solves  a long-standing puzzle known to general relativists. The integral on the right of \eq{ib1} leads to $(1/2)TA=2TS$ if we assume (for the sake of illustration) $T$= constant on the boundary and $S=A/4$. So, the Noether charge $Q$ is just the heat content  $Q=TS$, which is also obvious from \eq{flux3}. Therefore, the Noether charge is \textit{half} of the  equipartition energy of the surface $(1/2)TA=2TS$ when we attribute $(1/2)T$ per surface degree of freedom. For example, in the case of the Schwarzschild geometry,  the   equipartition energy of the surface is just the total mass $M=2TS$. \textit{But what the Noether charge measures is the heat content  $E-F=TS$ which has half this value, viz. $(M/2)$.} This leads to a ``problem''  in general relativity, when one tries to define the total mass of a spacetime (which behaves like the Schwarzschild spacetime asymptotically) using the (so-called) Komar integral. In this calculation, $\xi^a$ will be identified with the standard timelike Killing vector and the Noether potential will become the Komar potential. The integral one calculates with the Killing vector $\xi^a$ is identical to the computation of the Noether charge done above and one gets $(M/2)$. In standard general relativity, this was considered very puzzling because, in that context, we (at best!) only have a notion of energy but no notion of heat content ($TS$), free energy ($F=E-TS$), etc. The thermodynamic perspective --- which introduces the $\hbar$ through the definition of Davies-Unruh temperature $k_BT=(\hbar/c)(\kappa/2\pi)$) from an acceleration $\kappa$ --- tells us that the Noether charge is the heat content (enthalpy) $TS$ and \textit{not} the energy $2TS$, and that the result \textit{must} be $M/2$ for consistency of the formalism.

In short, in standard approach to GR, we can only interpret $M$ physically (as energy), while the thermodynamic approach allows us to \textit{also} interpret $M/2$ physically as the heat content $TS$. \textit{This is yet another case of the emergent paradigm giving us better insight into some puzzling features of standard general relativity.} 

Let us now move on to the main theme, viz. the dynamics of spacetime.  We 
take the dot product of the Noether current $J^a[\xi]$ in \eq{noe} (with $v^a=\xi^a$)
with $u_a$, use \eq{fin1}, introduce the gravitational dynamics through $R_{ab} = (8\pi L_P^2)\mathscr{F}_{ab} $ (where $\mathscr{F}_{ab}\equiv T_{ab}-(1/2)g_{ab}T)$ and integrate the result over a 3-dimensional region $\mathcal{R}$ with the measure $\sqrt{h}\, d^3x$. This gives:
\begin{equation}
 \int_\mathcal{R} \frac{d^3x}{8\pi L_P^2}  \, \sqrt{h}\, u_a g^{ij} \pounds_\xi N^a_{ij}
= \int_{\partial\mathcal{R}} \frac{d^2 x \, \sqrt{\sigma}}{L_P^2}\,\left( \frac{Na_\alpha r^\alpha}{4\pi}\right) - \int_\mathcal{R} d^3 x \, N \sqrt{h}\,( 2u^au^b \mathscr{F}_{ab})
\label{gravdyn}
\end{equation}
where $r_\alpha$ is the normal to the boundary of the 3-dimensional region.  We will again choose the boundary to be a $N(t,\mathbf{x})$= constant surface within the $t=$ constant surface.  
 We can then interpret, just as before,  $T_{\rm loc}=Na_\alpha r^\alpha/2\pi=Na/2\pi$ as the local Davies-Unruh temperature. Further, in
the second term, we see that $2N\mathscr{F}_{ab} u^a u^b = (\rho + 3p)N$ is the Komar energy density. Therefore  \eq{gravdyn} becomes:
\begin{equation}
 \frac{1}{8\pi L_P^2} \int_\mathcal{R} d^3x \sqrt{h}\, u_a g^{ij} \pounds_\xi N^a_{ij} =  \int_{\partial\mathcal{R}}\frac{d^2 x \, \sqrt{\sigma}}{L_P^2} 
\left( \half k_B T_{\rm loc}\right) - \int_\mathcal{R}d^3x\, \sqrt{h}\, \rho_{\rm Komar} 
\label{holoevl}
\end{equation} 

This result has a remarkable interpretation. When the spacetime is static and we choose the foliation such that $\xi^a$ is  identified with the  Killing vector, then $\pounds_\xi N^a_{ij} =0$ and the left-hand-side vanishes. The equality of the two terms on the right-hand-side represents the \textit{holographic equipartition} \cite{cqg04} if we define the bulk and surface degrees of freedom along the following lines: We count the number of surface degrees of freedom by allotting one `bit' for each Planck area:
\begin{equation}
 N_{\rm sur}\equiv\frac{A}{L_P^2}=\int_{\partial \mathcal{R}} \frac{\sqrt{\sigma}\, d^2 x}{L_P^2}
\end{equation} 
 The \textit{average} temperature $T_{\rm avg}$ of the boundary surface $\partial\mathcal{R}$ is given by:
 \begin{equation}
 T_{\rm avg}\equiv\frac{1}{A}\int_{\partial \mathcal{R}} \sqrt{\sigma}\, d^2 x\ T_{\rm loc}
\label{tav}
\end{equation} 
Finally, we  define the bulk degrees of freedom $N_{\rm bulk}$ as follows: \textit{If} the matter in the region $\mathcal{R}$ is in equipartition at the average surface temperature $T_{\rm avg}$, \textit{then} $|E| = (1/2) N_{\rm bulk} k_B T_{\rm avg}$; therefore, 
 the number of bulk degrees of freedom is:
\begin{equation}
N_{\rm bulk}\equiv \frac{|E|}{(1/2)k_BT_{\rm avg}}= \pm\frac{1}{(1/2)k_BT_{\rm avg}}\int_\mathcal{R} \sqrt{h}\ d^3x\; \rho_{\rm Komar}
\label{nbulkgen}
\end{equation} 
where $E$ is the total Komar energy in the bulk region $\mathcal{R}$ acting as a source of gravity.
(The $\pm$ sign is to ensure that $N_{\rm bulk}$ remains positive even when the  Komar energy turns negative.)  
This is the relevant value of $N_{\rm bulk}$ if the  equipartition holds for the  
 energy $E$ in the bulk region at the average surface temperature. 
Our result in \eq{holoevl} shows that \textit{comoving observers in any static spacetime} will indeed find:
\begin{equation}
 N_{\rm sur} = N_{\rm bulk} \qquad ({\rm Holographic \ equipartion})
 \label{key1}
\end{equation}
That is, the condition for holographic equipartition holds  in all static spacetimes.

More significantly, \eq{holoevl} shows clearly that \textit{it is the departure from holographic equipartition --- resulting in a non-zero value for the right-hand-side ---  drives the dynamical evolution of the spacetime.} Indeed, we can write \eq{holoevl} as:
\begin{equation}
\int \frac{d^3x}{8\pi L_P^2}\sqrt{h} u_a g^{ij} \pounds_\xi N^a_{ij} = \frac{1}{2} k_B T_{\rm avg} ( N_{\rm sur} - N_{\rm bulk})
\label{evlnsnb}
\end{equation} 
Even in a static spacetime, non-static observers  will perceive a departure from holographic equipartition because \eq{evlnsnb}  --- while being generally covariant --- is dependent on the foliation  through the normal $u_i$.  It is, of course,  possible for the same spacetime to be  described by two different class of observers
(i.e., foliations) such that the metric is static for one while it is non-static for the other. (A simple example is de Sitter spacetime which is static in spherically symmetric coordinates while being time dependent in FRW coordinates.) Unlike  Einstein's equation $G^a_b=8\pi T^a_b$, our \eq{evlnsnb} clearly distinguishes observers who perceive the spacetime to be static  (for which $N_{\rm sur} = N_{\rm bulk}$) from those who find it time dependent.

One can rewrite the left hand side of \eq{evlnsnb} by  relating $u_a g^{ij} \pounds_\xi N^a_{ij}$  to   more familiar constructs in the Hamiltonian formulation of relativity.
A direct computation  \cite{grtd} shows that  $\sqrt{h} u_a g^{ij} \pounds_\xi N^a_{ij}$ can be re-expressed as:
\begin{equation}
 \sqrt{h} u_a g^{ij} \pounds_\xi N^a_{ij}=-h_{ab}\pounds_\xi p^{ab}; 
\quad p^{ab}\equiv\sqrt{h}(Kh^{ab}-K^{ab})
\label{niskp}
\end{equation} 
allowing us to rewrite \eq{evlnsnb} in the form of \eq{start} presented at the beginning of the essay.

As I mentioned earlier, demanding the validity of \eq{start} or \eq{evlnsnb} for all foliations
 is mathematically equivalent to Einstein's equations.
 While \eq{evlnsnb} is a classical equation, individual parts of it (like $T_{\rm avg}, N_{\rm sur}$) contain  $\hbar$. This strengthens the idea that gravitational field equations have the same conceptual status as the equations of thermodynamics or fluid mechanics, with the Davies-Unruh temperature providing the link between microscopic and macroscopic descriptions of spacetime. It is remarkable that the dynamical evolution of the spacetime can be described in such an elegant, holographic language which is closer to thermodynamics than to geometry.


\begin{thebibliography}{99}


\bibitem{du} P C W. Davies, \textit{J. Phys.}, \textbf{A 8} (1975), 609;
W. G. Unruh, \textit{Phys. Rev.} \textbf{D 14}, 870 (1976).

\bibitem{grtd} T. Padmanabhan,	\textit{Gen.Rel.Grav.} \textbf{46} 1673 (2014) [arXiv:1312.3253]. 

 \bibitem{a8} T. Padmanabhan, \textit{Rept. Prog. Phys.}, \textbf{ 73} (2010) 046901, [arXiv:0911.5004]; \textit{J.Phys. Conf.Ser.}, \textbf{306}, 012001 (2011) [arXiv:1012.4476]; 
\textit{Res. Astro. Astrophys.}, \textbf{12}, 891 (2012) [arXiv:1207.0505]. 

\bibitem{KBP} K. Parattu, B. R. Majhi, T. Padmanabhan,  \textit{Phys.Rev.,} \textbf{D 87},  124011 (2013) [arXiv:1303.1535].


\bibitem{cqg04} T. Padmanabhan,\textit{Class.Quan.Grav.}, \textbf{21}, 4485 (2004) [gr-qc/0308070];  \textit{Mod.Phys.Lett.},  \textbf{A25} (2010) 1129, [arXiv:0912.3165];
\textit{ Phys.Rev.}, \textbf{D81} (2010) 124040, [arXiv:1003.5665].






\end{thebibliography}
\end{document}